\documentclass[useAMS,usenatbib]{mn2e}
\usepackage{amsfonts}
\usepackage{graphicx,float}
\usepackage[ansinew]{inputenc}
\usepackage{datetime}
\usepackage{titlesec}
\usepackage{subfig}
\title[Gauge Choice in Conformal Gravity]{Gauge Choice in Conformal Gravity}
\author[J. Sultana \& D. Kazanas]{
Joseph Sultana,$^{1}$\thanks{E-mail: joseph.sultana@um.edu.mt}
Demosthenes Kazanas,$^{2}$\thanks{E-mail: Demos.Kazanas-1@nasa.gov}
\\
$^{1}$Department of Mathematics, Faculty of Science, University of Malta, Msida MSD 2080, Malta\\
$^{2}$ Astrophysics Science Division, NASA/Goddard Space Flight Center, Greenbelt, Maryland 20771, USA
}
\begin{document}

\date{}

\pagerange{\pageref{firstpage}--\pageref{lastpage}} \pubyear{2016}

\maketitle

\label{firstpage}

\begin{abstract}
In a recent paper (MNRAS 458, 4122 (2016)) K. Horne examined the effect of a
conformally coupled {scalar field (referred to as Higgs field)} on the
{Mannheim-Kazanas metric $g_{\mu\nu}$, i.e. the static spherically
symmetric metric within the context of conformal gravity (CG), and studied its effect
on the rotation curves of galaxies}. He showed that for a Higgs field {of the
form} $S(r) = S_0 a/(r + a)$, where $a$ is a radial length scale, the equivalent
Higgs-frame Mannheim-Kazanas metric $\tilde{g}_{\mu\nu} = \Omega^2 g_{\mu\nu}$, with $\Omega =
S(r)/S_{0}$, lacks the linear $\gamma r$ term, which {has been employed in}
the fitting of the galactic rotation curves {without the need to invoke dark
matter}. In this brief note we point out that the representation of the Mannheim-Kazanas metric in
a gauge where it lacks the linear term has already been presented by others,
including Mannheim and Kazanas themselves, without the need to introduce a
conformally coupled Higgs field. {Furthermore, Horne argues that the absence
of the linear term resolves the issue of light bending in the wrong direction, i.e.
away from the gravitating mass, if $\gamma r > 0$ in the Mannheim-Kazanas metric, a condition
necessary to resolve the galactic dynamics in the absence of dark matter.} 
In this case we also point out that the elimination of the linear term is not even
required because the sign of the $\gamma r$ term in the metric can be easily reversed
by a simple gauge transformation, {and also that the effects of this term are
indeed too small to be observed}.
\end{abstract}


\section{Introduction}
Conformal (Weyl) gravity adopts the principle of local conformal invariance of
spacetime under local conformal stretching $g_{\mu\nu}(x) \rightarrow \Omega^2(x)
g_{\mu\nu}(x)$, where $\Omega(x)$ is a smooth strictly positive function. This is used
as the supplementary condition that fixes the gravitational action, instead of
requiring that the theory be no higher than second order as in the case of the
Einstein-Hilbert action. This restrictive conformal invariance leads to a fourth order
theory with the unique conformally invariant action
\begin{eqnarray} I_{W} & = & -\frac{1}{2\alpha}\int
d^4x(-g)^{1/2}C_{\lambda\mu\nu\kappa}C^{\lambda\mu\nu\kappa} \nonumber\\
      & = & -\frac{1}{\alpha}\int d^4x(-g)^{1/2}[R_{\mu\kappa}R^{\mu\kappa} -
      (R^{\nu}_{\nu})^2/3]\\
      & ~~~~& +{\rm a~total~derivative}, \label{action}
\end{eqnarray}
where $C_{\lambda\mu\nu\kappa}$ is the conformal Weyl tensor and
$\alpha$ is a purely dimensionless coefficient. Varying the
action in (\ref{action}) with respect to the metric leads to the field equations
\begin{equation}
\frac{1}{\sqrt{-g}}g_{\mu\alpha}g_{\nu\beta}\frac{\delta
I_{W}}{\delta g_{\alpha\beta}} =-\frac{1}{\alpha} W_{\mu\nu},
\label{weyl_field_eqns}
\end{equation}
where
\begin{equation}
W_{\mu\nu} = 2C^{\alpha\ \ \beta}_{\ \mu\nu\ ;\beta\alpha} +
C^{\alpha\ \ \beta}_{\ \mu\nu}R_{\alpha\beta}. \label{w1}
\end{equation}
In the presence of sources the full field equations are obtained by variation with
respect to the metric of the total action $I = I_{W} + I_{M}$, where $I_{M}$ is the
action corresponding to the source. This gives
\begin{equation}
W_{\mu\nu} = \frac{\alpha}{2}T_{\mu\nu},
\end{equation}
where $T_{\mu\nu} = 2(-g)^{-1/2}\delta I_M/\delta g^{\mu\nu}$ is a conformally
invariant stress-energy tensor. It can be shown \citep{conformal1} that the tensor
$W_{\mu\nu}$ vanishes when $R_{\mu\nu}$ is zero, so that any vacuum solution of
Einstein's field equations is also a vacuum solution of Weyl gravity, with the converse
not being necessarily true. Despite the highly nonlinear character of the field
equations, a number of exact solutions
\citep{conformal1,furthersolutions1,furthersolutions2,furthersolutions3,furthersolutions4,cylindrical1,cylindrical2}
of conformal Weyl gravity have been found. The  Mannheim-Kazanas (MK) metric is the spherically symmetric
vacuum solution in conformal gravity, satisfying $W_{\mu\nu}=0$ and describing the
geometry outside a spherical body. This is given, up to a conformal factor, by the
metric \citep{conformal1}
\begin{equation}
ds^2 = -B(r)dt^2 + \frac{dr^2}{B(r)} + r^2(d\theta^2 + \sin^2\theta
d\phi^2), \label{generalmetric}
\end{equation}
where
\begin{equation}
B(r) = 1 - \frac{\beta(2 - 3\beta\gamma)}{r} - 3\beta\gamma + \gamma
r - k r^2, \label{eq:metric}
\end{equation}
and $\beta,\ \gamma,\  \mbox{and}\ k$ are integration constants. This solution includes
the Schwarzschild $(\gamma = k = 0)$ and the Schwarzschild-de Sitter $(\gamma = 0)$
solutions as special cases, with the latter requiring the presence of a cosmological
constant in Einstein gravity. The constant $\gamma$, whose origin remains
uncertain, has dimensions of acceleration; as such, {conformal gravity is the unique (to our
knowledge) gravitational theory that provides a solution with a characteristic},
constant, Rindler-like acceleration, without the need to introduce one in the
Lagrangian (such as the relativistic implementation of MOND, TeVeS \citep{bekenstein}).
{Its magnitude is heuristically estimated to be $c H_0$, on the basis of the
asymptotically non-flat form of the resulting metric, a value consistent with the most
recent observations of dynamics of galaxies \citep{McGaugh}}. An important consequence of
this linear term in the metric which generated a growing interest in conformal gravity,
has been the fitting of galactic rotational curves
\citep{conformal1,mannheim93,mannheim97,mannheim11,obrien12} without the need to assume
dark matter as in Einstein's gravitational theory. This is achieved \citep{conformal1}
by associating $\gamma$ with the inverse Hubble length, i.e. $\gamma \simeq 1/R_H$,
such that the effects of this acceleration are comparable to those due to the Newtonian
potential term $2\beta/r \equiv r_s/r$ ($r_s$ is the Schwarzschild radius), on length
scales given by
\begin{equation}
r_s / r^2 \simeq \gamma \simeq 1/R_H ~~{\rm or}~~ r \simeq (r_s \,
R_H)^{1/2}. \label{eq:MRrelation}
\end{equation}
For example in the case of a galaxy of mass $M \simeq 10^{11}\;
{\rm M}_{\odot}$ with $r_s \simeq 10^{16}$ cm and $R_H \simeq
10^{28}$ cm, this scale is $r \sim 10^{22}$ cm, i.e. roughly the
size of the galaxy.  Eq. (\ref{eq:MRrelation}) doesn't
fix a particular length scale but represents a continuum of scales at which
the contribution from the linear term becomes significant.
Objects along this sequence do not include only galaxies but, at
larger scales also galaxy clusters and at lower scales globular
clusters, only recently found to require the presence of dark matter
in order to account for the observed dynamics \citep{scarpa,kazanas08}. Besides the fitting of galactic rotational curves the effect of the linear $\gamma r$ term in the MK-metric (\ref{eq:metric}) on the classical tests, such as the bending
of light \citep{deflection1,amore06,sultana10} and perihelion precession \citep{sultana12} have been investigated.

In the next section we discuss the results of \citet{horne16} and point out that the presence of a conformally coupled scalar field to the theory is not required to eliminate the linear term from the MK-metric. Moreover in section 3 we show that for light bending one can still achieve a positive deflection from the linear term by choosing an appropriate gauge.  We summarize our results in the conclusion.

\section{The choice of gauge}

\cite{brihaye09} obtained an exact static spherically symmetric solution to scalar tensor conformal gravity where the gravitational Lagrangian is given by
\begin{equation}
L_{g} = \frac{1}{\alpha}\left(-\frac{1}{2}C_{\kappa\lambda\mu\nu}C^{\kappa\lambda\mu\nu} - \frac{1}{2}\nabla_{\lambda}S\nabla^{\lambda}S - \frac{1}{12}RS^2 - \frac{\nu}{4}S^4\right),
\end{equation}
with $S$ being the conformally coupled scalar field satisfying the scalar field equation
\begin{equation}
\nabla_{\mu}\nabla^{\mu}S - \nu S^3 - \frac{RS}{6} = 0 \label{sfeq},
\end{equation}
and $\nu$ is the self-coupling parameter. In this case the field equations are given by
\begin{equation}
W_{\mu\nu} = \frac{1}{2}S_{\mu\nu}, \label{ccfeq}
\end{equation}
where
\begin{equation}
S_{\mu\nu} = \nabla_{\mu}S\nabla_{\nu}S - \frac{1}{2}g_{\mu\nu}\nabla_{\lambda}S\nabla^{\lambda}S
- \frac{\nu}{4}S^4g_{\mu\nu} +\frac{1}{6}(g_{\mu\nu}\nabla^{\lambda}\nabla_{\lambda}S^2
- \nabla_{\mu}\nabla_{\nu}S^2 + G_{\mu\nu}S^2);
\end{equation}
$G_{\mu\nu}$ being the Einstein tensor. The spherically symmetric exact solution to
(\ref{sfeq}) and (\ref{ccfeq}) obtained by \cite{brihaye09} is given by
\begin{equation}
ds^2 = -B(r)dt^2 + \frac{dr^2}{B(r)} + r^2 (d\theta^2 + \sin^2\theta d\phi^2),
\end{equation}
where
\begin{equation}
B(r) = (1 + r/a)^2 - \frac{r_h}{r}\frac{(1 + r/a)^3}{(1 + r_h/a)} +
\frac{\nu S_0^2 r_h^2}{2}\left(\frac{r^2}{r_h^2} -
\frac{r_h}{r}\frac{(1 + r/a)^3}{(1 + r_h/a)^3}\right) \label{brihaye},
\end{equation}
and
\begin{equation}
S(r) = \frac{S_0}{1 + r/a} \label{sfbrihaye},
\end{equation}
where $r_h$, $a$ and $S_0$ are free parameters. This solution is not a generalization
of the MK-solution {but it} is equivalent to it. In fact one can easily check
that the energy momentum tensor $S_{\mu\nu}$ on the RHS of (\ref{ccfeq}) vanishes for
the above scalar field and metric. One can also relate the free parameters in
(\ref{brihaye}) with those in the MK-solution by
\begin{equation}
\beta = \frac{2ar_h(a + r_h)^2 + a^3r_h^3S_0^2\nu}{2(2a - r_h)(a + r_h)^2 - 3a^2r_h^3S_0^2\nu},
\end{equation}
\begin{equation}
\gamma = -\frac{1}{a} + \frac{3}{a + r_h} - \frac{3ar_h^3S_0^2\nu}{2(a+r_h)^3},
\end{equation}
and
\begin{equation}
k = -\frac{1}{a(a+r_h)} - \frac{a(a^2 + 3ar_h + 3r_h^2)S_0^2\nu}{2(a+r_h)^3}.
\end{equation}
In his paper, \cite{horne16} examined the effect of the conformally coupled scalar
field $S(r)$ in (\ref{sfbrihaye}) on the rotational curves of galaxies. In particular
he showed that when the MK-metric written in the form (\ref{brihaye}) is stretched to
what he calls the ``Higgs frame'', which is given by
\begin{equation}
g_{\mu\nu} \rightarrow \tilde{g}_{\mu\nu} = \Omega^2g_{\mu\nu} = \left(\frac{S}{S_0}\right)^2g_{\mu,\nu},
\end{equation}
the corresponding metric $\tilde{g}_{\mu\nu}$ lacks the linear term $\gamma\tilde{r}$
in the potential $\tilde{g}_{00}$, and therefore this has a significant effect on the
shape of the resulting rotation curve in the Higgs frame geometry. Now the fact that
the linear term in the MK-metric can be eliminated by conformal transformation has been
known for a long time (see for example \cite{furthersolutions1,schmidt00}) and the
presence of a conformally coupled scalar field doesn't play any role in this. In fact
as already remarked above and in \cite{brihaye09} the scalar field in (\ref{sfbrihaye})
is trivial in the sense that it has no effect on the geometry because the corresponding
energy momentum tensor $S_{\mu\nu}$ on the RHS of the field equations (\ref{ccfeq})
vanishes. So for example starting with the MK-metric in (\ref{eq:metric}) and
redefining the radial coordinate by
\begin{equation}
r = \frac{\tilde{r}}{1 + \tilde{\alpha}\tilde{r}}, \label{rescaling}
\end{equation}
\noindent the conformally transformed metric $\tilde{g}_{\mu\nu} = (1 +
\tilde{\alpha}\tilde{r})^2 g_{\mu\nu}$, with $\tilde{\alpha} = \gamma/(3\beta\gamma -
2)$ takes the Schwarzschild de-Sitter form
\begin{equation}
ds^2 = -\tilde{B}(\tilde{r})dt^2 + \frac{d\tilde{r}}{\tilde{B}(\tilde{r})} + \tilde{r}^2(d\theta^2 + \sin^2\theta d\phi^2),
\end{equation}
with
\begin{equation}
\tilde{B}(\tilde{r}) = 1 - \frac{(2 - 3\beta\gamma)\beta}{\tilde{r}} - \tilde{r}^2\frac{\gamma^2 - \beta\gamma^3 + k(2 - 3\beta\gamma)^2}{(2 - 3\beta\gamma)^2}
\end{equation}
{This expression now reflects the metric in Horne's Higgs-field frame, i.e. a
metric conformally equivalent to that of MK by the $r$ re-scaling given in (\ref{rescaling}).}

\section{Gravitational Lensing}

{In further discussion in his paper, \cite{horne16} argues in favor of the
metric in the Higgs conformal gauge vis-\'a-vis gravitational lensing. It has been
noted that the $\gamma r$ term in the MK metric produces a negative angle light
bending, i.e. away from the gravitating source  and proportional to the photon impact
parameter, in gross disagreement with observation. He therefore argues that elimination
of this term will reduce gravitational lensing to the standard treatment without the
unwanted behavior.}
%
One has to keep in mind that unlike the case of the galactic rotational curves the
effect of the linear term on the bending of light in conformal gravity is negligible
and, as expected, decreases with the impact parameter. {This is not surprising
given that the metric obtained setting $\beta =0$ in the MK solution is conformally
flat and therefore it should leave invariant the null geodesics}. In fact, as shown in
\cite{sultana10},  the contribution to the bending angle from the linear term in the
MK-metric is $-2\beta^2\gamma/b$, where $b$ is the impact parameter, and so the ratio
of this angle and the conventional bending angle $4\beta/b$ arising from the $1/r$ term
in the metric, is of the order of $\beta\gamma$, which for a galaxy is just
$\beta\gamma \sim 10^{-12}$, i.e. insignificant for practical purposes. This contrasts
with earlier claims \citep{deflection1,pireaux04a,pireaux04b,amore06} that this
contribution from the linear term is $-\gamma b$, which becomes significant for larger
values of the impact parameter $b$. These claims were based on calculations done with the improper
use of the bending angle formula in spaces that are not asymptotically flat. Also apart
from the negligible magnitude of the term containing $\gamma$ in the expression for the
bending angle, the issue of its (negative) sign should not be considered undesirable,
given that the true nature of $\gamma$ in the metric is still unknown. In fact for
large $r$ the MK-metric in (\ref{eq:metric}) reduces \citep{conformal1} to a metric
which is conformal to the Robertson-Walker metric with an arbitrary scale factor and
with a spatial curvature that depends on $k$ and $\gamma$, which means that this
parameter knows about the background spacetime and is not strictly system dependent. In
other words by exploiting the conformal structure of the theory, the linear term in the
metric facilitates the interpolation of the solution between the Schwarzschild and
Robertson-Walker geometries in a continuous and smooth manner.  The cosmological
constant $\Lambda$ (which is equivalent to $k$) in the Schwarzschild de-Sitter geometry
produces \citep{rindler07} a negative contribution to the bending angle. But even if
one questions the validity of the cosmological nature of the parameter $\gamma$, one
can still make the theory attractive to null geodesics through an appropriate choice of
the conformal gauge \citep{edery01} without the need to take $\gamma < 0$ as stated in
\cite{horne16}, and so one can still do gravitational lensing in the MK-frame. Hence one
can start with the metric in (\ref{eq:metric}) and apply the conformal rescaling $ds^2
\rightarrow \tilde{\Omega}ds^2$ followed by the coordinate transformation
\begin{equation}
r' = r\sqrt{\tilde{\Omega}},
\end{equation}
and assume that the conformally transformed metric has the form
\begin{equation}
ds^2 = -B'(r')dt^2 + \frac{dr'^2}{B'(r')} + r'^2 (d\theta^2 +\sin^2\theta d\phi^2),
\end{equation}
where
\begin{equation}
B'(r') = 1 - \frac{\beta(2 - 3\beta\gamma)}{r'} - 3\beta\gamma - \gamma r' - k r'^2, \label{eq:metricve}
\end{equation}
contains a negative linear term. Then in a weak field approximation $\beta/r << 1 ,
\beta\gamma <<1$, and keeping only first order terms, the necessary conformal factor is
given by $\tilde{\Omega} = 1 - 2\gamma r$. The negative linear term in
(\ref{eq:metricve}) will then yield a contribution to the bending angle in the
direction of the object, when $\gamma > 0$.

\section{Conclusion}

In this paper we have shown that one can do a simple gauge choice to transform the
MK-metric in a form that does not contain the linear term and in this sense the use of
a conformally coupled Higgs field to the theory and the associated Higgs frame is not
required. The field itself is trivial since it yields a vanishing energy momentum
tensor on the RHS of the field equations. We also point out that the sign of the linear
term in the MK-metric can be changed by a similar gauge transformation and since null
geodesics are preserved under conformal transformations, one can use such a gauge to
obtain a positive contribution for the deflection angle, without actually eliminating
it.

\section*{Acknowledgments}
J.S. gratefully acknowledges financial support from the University of Malta during his
visit at NASA-GSFC and the hospitality of the Astrophysics Science Division of GSFC.

\bsp

\label{lastpage}

\end{document}